\documentclass[conference]{IEEEtran}
\IEEEoverridecommandlockouts

\usepackage{cite}
\usepackage{amsmath,amssymb,amsfonts}
\usepackage{algorithmic}
\usepackage{graphicx}
\usepackage{textcomp}
\usepackage{xcolor}
\def\BibTeX{{\rm B\kern-.05em{\sc i\kern-.025em b}\kern-.08em
    T\kern-.1667em\lower.7ex\hbox{E}\kern-.125emX}}
\begin{document}

\title{Brian Intensify: An Adaptive Machine Learning Framework for Auditory EEG Stimulation and Cognitive Enhancement in FXS\\
}

\author{\IEEEauthorblockN{Zag ElSayed} 
\IEEEauthorblockA{\textit{School of Information Technology} \\
\textit{University of Cincinnati} \\
Ohio, USA} \\
\and
\IEEEauthorblockN{Grace Westerkamp, Jack Yanchen Liu, Ernest Pedapati}
\IEEEauthorblockA{\textit{Division of Child and Adolescent Psychiatry} \\
\textit{Cincinnati Children’s Hospital Medical Center}\\
Ohio, USA} \\
}

\maketitle

\begin{abstract}
Neurodevelopmental disorders such as Fragile X Syndrome (FXS) and Autism Spectrum Disorder (ASD) are characterized by disrupted cortical oscillatory activity, particularly in the alpha and gamma frequency bands. These abnormalities are linked to deficits in attention, sensory processing, and cognitive function. In this work, we present an adaptive machine learning-based brain-computer interface (BCI) system designed to modulate neural oscillations through frequency-specific auditory stimulation to enhance cognitive readiness in individuals with FXS. EEG data were recorded from 38 participants using a 128-channel system under a stimulation paradigm consisting of a 30-second baseline (no stimulus) followed by 60-second auditory entrainment episodes at 7Hz, 9Hz, 11Hz, and 13Hz. A comprehensive analysis of power spectral features (Alpha, Gamma, Delta, Theta, Beta) and cross-frequency coupling metrics (Alpha-Gamma, Alpha-Beta, etc.) was conducted. The results identified Peak Alpha Power, Peak Gamma Power, and Alpha Power per second per channel as the most discriminative biomarkers. The 13Hz stimulation condition consistently elicited a significant increase in Alpha activity and suppression of Gamma activity, aligning with our optimization objective. A supervised machine learning framework was developed to predict EEG responses and dynamically adjust stimulation parameters, enabling real-time, subject-specific adaptation. This work establishes a novel EEG-driven optimization framework for cognitive neuromodulation, providing a foundational model for next-generation AI-integrated BCI systems aimed at personalized neurorehabilitation in FXS and related disorders.
\end{abstract}


\section{Introduction}
Several neurodevelopmental disorders such as Fragile X Syndrome (FXS) and Autism Spectrum Disorder (ASD) are characterized by atypical patterns of brain connectivity and oscillatory dysfunction, which contribute to deficits in learning, sensory processing, and executive function~\cite{hagerman2017fragile,schmitt2022altered}. In particular, abnormalities in alpha (8–12 Hz) and gamma (30–100 Hz) oscillations have been consistently observed in FXS, where reduced alpha power and elevated gamma activity are associated with poor cognitive outcomes and heightened sensory hypersensitivity~\cite{ethridge2017neural, takarae2024eeg, proteau2024specific}. Neural oscillations play a vital role in synchronizing brain activity across various networks~\cite{palva2011functional}, so selectively modulating these rhythms could be highly beneficial for therapeutic treatments.

Electroencephalography (EEG)-based analysis of brain rhythms provides a non-invasive, real-time window into neural dynamics, offering an ideal platform for brain-computer interface (BCI) applications aimed at neuromodulation~\cite{makeig2002dynamic}. Recent progress in neurostimulation linked to auditory cues has shown that rhythmic sound stimuli can synchronize neural oscillations, effectively guiding brain activity toward more optimal states through a process referred to as neural entrainment.~\cite{gourevitch2020oscillations,lakatos2007neuronal}.

However, conventional auditory stimulation protocols often rely on static, one-size-fits-all paradigms, ignoring individual variability in neural responsiveness. Such limitations highlight the need for adaptive systems that dynamically tailor stimulation parameters based on continuous EEG feedback.

Machine learning (ML) provides powerful tools for modeling complex brain-stimulation relationships and enabling real-time adaptation based on subject-specific neural signatures ~\cite{makeig2002dynamic}. In particular, supervised learning models can be trained to predict the optimal stimulation parameters that maximize favorable EEG biomarkers, such as enhancing alpha power and suppressing gamma activity, which are associated with improved cognitive readiness and sensory regulation ~\cite{voytek2015age}.

\begin{figure}[htbp]
\centerline{\includegraphics[width =\linewidth]{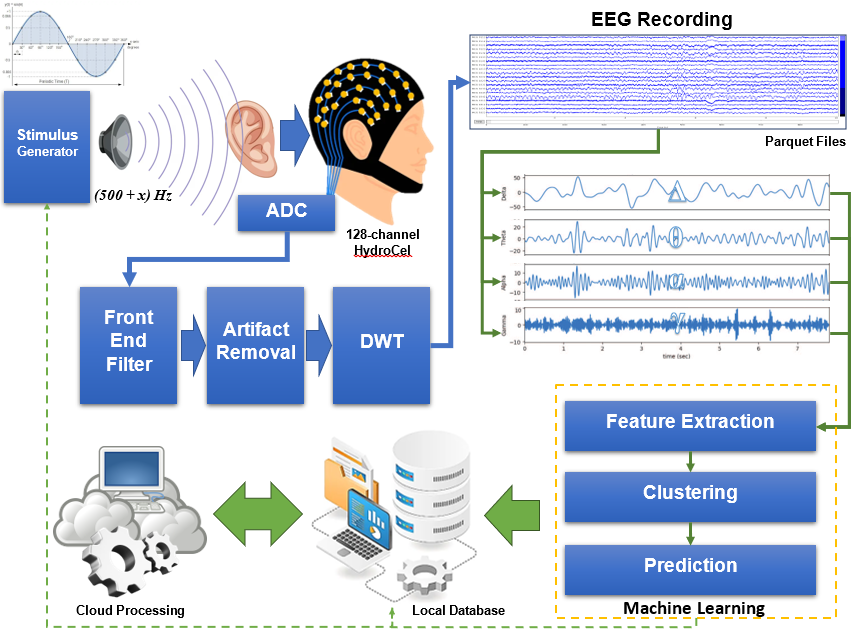}}
\caption{The proposed architecture diagram with data paths.}
\label{fig1}
\end{figure}

In this study, we present an adaptive ML based BCI framework that leverages EEG biomarkers to optimize auditory stimulation in individuals with FXS. Specifically, we investigated the effects of auditory stimuli at distinct frequencies (7Hz, 9Hz, 11Hz, and 13Hz) on EEG responses collected from 38 participants using a 128-channel EEG system. We extracted key neurophysiological features, including Peak Alpha Power, Peak Gamma Power, and cross-frequency coupling (CFC) metrics, such as Alpha-Gamma coupling, to identify the most responsive biomarkers. Our findings reveal that 13Hz auditory stimulation significantly enhances alpha activity while simultaneously reducing gamma oscillations, establishing a foundation for real-time adaptive neurostimulation, shown in Fig.\ref{fig1}.

This paper makes the following key contributions: First, it proposes an EEG-driven adaptive BCI system utilizing machine learning for the real-time optimization of auditory stimulation. Second, identify Peak Alpha Power, Peak Gamma Power, and Alpha per second per channel as robust biomarkers for cognitive enhancement in FXS. Third, it demonstrates that 13Hz auditory stimulation produces the most favorable neuromodulatory effects, suggesting a targeted intervention strategy. Finally, it establishes a scalable framework for next-generation AI-integrated neurorehabilitation platforms applicable to FXS and potentially broader neurodevelopmental conditions.

\section{Background}

\subsection{Neural Oscillations and EEG Biomarkers}
Neural oscillations serve as fundamental mechanisms for communication and coordination within and between cortical networks. Among these, alpha (7–13.5 Hz) and gamma (30–100 Hz) frequency bands play critical roles in modulating attention, sensory processing, and cognitive control ~\cite{palva2007new} ~\cite{canolty2010functional}. Alpha activity has traditionally been associated with cortical inhibition, sensory gating, and top-down attentional modulation, while gamma activity reflects local network excitation, binding processes, and bottom-up sensory input [3], [4]. 

The balance between these bands, particularly the Alpha-Gamma interaction, is thought to govern the efficiency of neural information processing. While popularly assumed that Delta waves (0.5-4 Hz) are associated with deep sleep and unconscious processes. Theta waves (4-8 Hz) with light sleep, REM, creativity, and meditation. Alpha waves (7-13.5 Hz)with relaxed wakefulness and calm alertness. Beta waves (13.5-30 Hz) are associated with active thinking and focused attention. Gamma waves (30-100 Hz): with high-level cognitive processing. However, this study focuses on the echoed information from the resulting waveforms as a corresponding output of brain mechanism activation, such as learning.

\begin{figure}[htbp]
\centerline{\includegraphics[width =\linewidth]{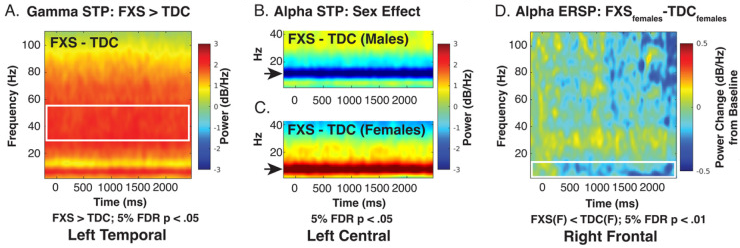}}
\caption{Single-trial power (STP) and event-related spectral perturbation (ERSP) in response to auditory chirp~\cite{pedapati2025frontal}.}
\label{fig2}
\end{figure}

In Fragile X Syndrome (FXS), studies have consistently shown reduced alpha power and elevated gamma power, often referred to as "gamma hypersynchrony"~\cite{wang2013resting}~\cite{rojas2014gamma}~\cite{goodspeed2023electroencephalographic}. These dysregulations have been linked to excessive neural excitability and sensory hypersensitivity, hallmark features of FXS and related Autism Spectrum Disorders (ASDs)~\cite{razak2021neural}. Furthermore, disruptions in cross-frequency coupling (CFC)—especially phase-amplitude coupling (PAC) between slower (e.g., alpha) and faster (e.g., gamma) rhythms have been proposed as biomarkers for atypical neurodevelopment~\cite{pedapati2025frontal}, where the authors show power spectrum in response to auditory chirp, shown in Fig.\ref{fig2}.

Electroencephalography (EEG), due to its high temporal resolution and non-invasive nature of the signal recording sensors, it has been widely adopted to measure these biomarkers. The spatiotemporal properties of EEG signals allow researchers to track instantaneous changes in brain activity, which is particularly useful for brain-computer interfaces (BCIs) that aim to deliver feedback-based stimulation. Importantly, EEG can quantify not only spectral power per channel and timepoint but also the dynamic coupling between frequencies and regions, providing a multidimensional signal space for optimization~\cite{makeig2002dynamic}.

\subsection{Neural Modulation via Auditory Stimulation}
Auditory stimulation, especially rhythmic and frequency-specific sound delivery, has been shown to exert powerful neuromodulatory effects on cortical oscillations through a mechanism known as neural entrainment. In this process, external rhythmic stimuli align the phase and amplitude of endogenous neural rhythms with the stimulus frequency, effectively shifting brain activity toward more functional states~\cite{obleser2019neural} ~\cite{lakatos2008entrainment}. This principle has been successfully applied in domains ranging from attention enhancement to memory consolidation and neurorehabilitation ~\cite{rajendran2019frequency}.

Auditory entrainment predominantly operates through auditory steady-state responses (ASSRs), which reflect the brain’s natural tendency to synchronize with periodic acoustic inputs. When the auditory cortex receives continuous rhythmic input, particularly in the alpha or gamma frequency ranges, the EEG shows measurable spectral peaks at the stimulus frequency, indicating entrainment~\cite{henao2020entrainment}. The strength and stability of ASSRs are linked to cognitive function, and deviations from typical entrainment patterns are frequently observed in neurodevelopmental disorders, including FXS and ASD ~\cite{kenny2022eeg}.

Furthermore, auditory stimulation at specific frequencies has been shown to modulate cross-frequency coupling (CFC)—a critical mechanism for information integration across neural scales. For instance, alpha-band entrainment has been associated with enhanced alpha-gamma phase-amplitude coupling, facilitating more efficient top-down control and sensory filtering~\cite{effenberger2025functional}. Conversely, unregulated gamma activity, as commonly observed in FXS, is linked to poor CFC and excessive neural noise~\cite{canolty2010functional}.

Empirical studies support the use of alpha-range auditory stimulation (10–13 Hz) to improve cognitive readiness. For example, 10Hz rhythmic tones have been reported to increase attention and reduce anxiety, stimulation in the lower theta range (e.g., 4–7 Hz) has been linked to relaxation and drowsiness, which may not be conducive to cognitive activation in certain populations~\cite{han2018increased}.

Importantly, the efficacy of auditory stimulation is highly individualized and state-dependent, underscoring the need for adaptive systems that can monitor EEG responses in real-time and dynamically adjust stimulation parameters. Our system builds on this principle by employing machine learning to identify the most effective auditory stimulation frequency per subject, thereby optimizing cognitive biomarkers such as increased alpha power and reduced gamma overactivity.

\subsection{Related Work in EEG-based Neurostimulation and Adaptive BCI}
BCI for neuromodulation has advanced significantly in recent years, driven by developments in closed-loop signal processing, machine learning, and real-time EEG decoding. These systems are increasingly used in cognitive rehabilitation, neurofeedback, and sensory integration therapies. One promising domain is auditory-driven neurostimulation, wherein external rhythmic auditory stimuli are used to entrain internal brain rhythms~\cite{bauer2021rhythmic}. Such entrainment can occur via mechanisms of auditory steady-state responses (ASSRs), wherein the brain synchronizes to periodic auditory inputs, potentially resetting dysfunctional oscillatory dynamics.

Several studies have demonstrated that specific auditory frequencies can modulate cortical activity. For instance, 10Hz auditory tones have been shown to increase alpha power and promote relaxation. In contrast, 40Hz auditory stimulation has been investigated for its potential to restore gamma-related connectivity in Alzheimer’s disease ~\cite{clements2016short}. However, most current systems use fixed-frequency paradigms and fail to adapt to the dynamic and individualized nature of EEG responses—especially in neurodivergent populations such as FXS and ASD.

Recently, another study introduced an auditory neurofeedback platform where spectral EEG features were used to regulate ambient sound environments for individuals with sensory sensitivities~\cite{ramirez2015musical} and implemented a deep reinforcement learning agent to modulate auditory stimulation in a neurofeedback loop, demonstrating improvements in sustained attention and working memory in neurotypical adults~\cite{fedotchev2023methods}. While these systems highlight the promise of real-time adaptive neurostimulation, they often rely on low-density EEG setups and do not target biomarker-specific frequency bands such as alpha-gamma dysregulation—key in FXS and ASD. Furthermore, few systems combine high-density EEG, biomarker-driven target optimization, and frequency-specific auditory entrainment. In contrast, our approach leverages per-second spectral and cross-frequency EEG features to guide machine learning-driven stimulation selection, tailored specifically to address the oscillatory phenotype of FXS using individualized neural signatures.

Our work builds upon these advances by introducing a biomarker-driven, ML-optimized auditory stimulation framework. It is unique in several respects:
\begin{itemize}
    \item Targets fragile X-specific EEG phenotypes (e.g., Alpha, Gamma, CFC disruption). 
    \item Applies frequency-specific auditory stimuli and evaluates multiple entrainment frequencies (7Hz, 9Hz, 11Hz, 13Hz).
    \item It integrates real-time adaptive ML models to select optimal stimulation frequencies per individual.
    \item It leverages per-second EEG features across 128 channels, enabling high-resolution dynamic modeling of brain responses for a future source allocation.
\end{itemize}

By combining EEG analytics, auditory entrainment, and intelligent optimization, our approach aims to push the frontier of personalized BCI-based neuro therapies for neurodevelopmental disorders.

\section{Methods}
\subsection{Participants and Clinical Protocol}
The study was conducted with 38 human participants (mean age: 12.4 ± 3.1 years) diagnosed with Fragile X Syndrome (FXS). All participants were clinically evaluated and recruited through CCHMC. Informed consent was obtained from all participants or legal guardians. Participants were seated in a quiet, electromagnetically shielded room and instructed to remain still with eyes open while auditory stimuli were delivered binaurally through calibrated headphones. 

The experimental setup was designed to elicit reliable auditory EEG responses in a controlled, distraction-free environment. Participants were seated in a comfortable chair inside a soundproof, electromagnetically shielded chamber. They were instructed to remain still, maintain a neutral gaze, and minimize movement during the EEG recording. Auditory stimuli were presented binaurally through calibrated high-fidelity headphones to ensure consistent sound delivery across trials.

\subsection{Auditory Stimulation Paradigm}
\begin{figure}[htbp]
\centerline{\includegraphics[width =\linewidth]{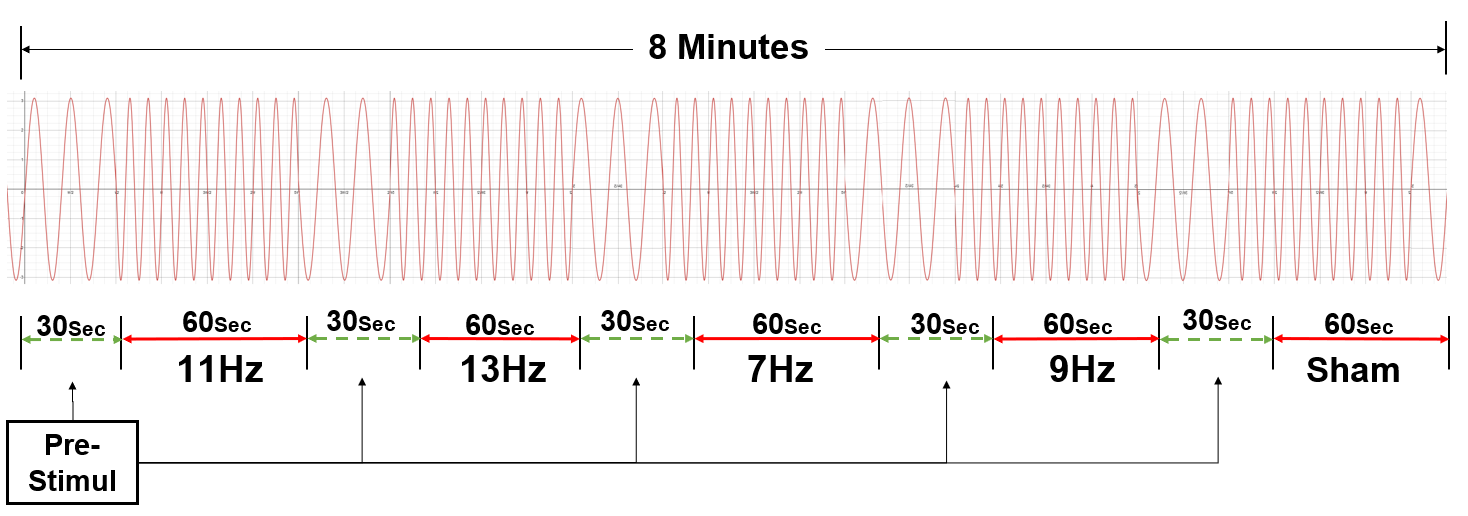}}
\caption{Auditory stimulation input file structure.}
\label{fig4}
\end{figure}

The stimulation paradigm was designed to study oscillatory entrainment and consisted of a 30-second resting-state baseline, followed by four 60-second frequency-specific auditory stimulation blocks delivered in randomized order at 7Hz, 9Hz, 11Hz, and 13Hz. Each block consisted of rhythmic click trains embedded in sine noise, maintaining consistent amplitude and volume across trials. The structure of the auditory impute file is shown in Fig.\ref{fig4}.

The use of relatively low frequencies was informed by prior research on auditory steady-state responses (ASSRs) and their alignment with EEG rhythms critical to cognitive processing in children. This frequency-specific design allows for the investigation of EEG biomarkers sensitive to subtle changes in external rhythmic drive, a key requirement for later adaptive modeling.

\subsection{EEG Acquisition and Pre-processing}
EEG signals were recorded using a 128-channel HydroCel electrode net system, sampling at 1000 Hz. Participants were seated in a soundproof, electro-magnetically shielded chamber to minimize environmental noise and artifacts, and the data quality was monitored in real-time.

Raw EEG data were preprocessed using EEGLAB (MATLAB) and MNE-Python to ensure high-quality feature extraction. A bandpass filter of 0.5–100 Hz was applied, followed by notch filtering at 57-63 Hz to eliminate powerline noise.

Independent Component Analysis (ICA) was then employed to remove non-neural artifacts such as ocular movements and muscle activity. The cleaned EEG data were segmented into non-overlapping 1-second epochs, time-locked to auditory stimulus onsets. This epoching resolution enabled fine-grained temporal modeling of dynamic spectral changes, which is essential for real-time machine learning prediction.

\subsection{Feature Extraction and Biomarker Selection}
From each 1-second EEG epoch, we extracted a rich set of spectral and coupling-based features that capture key neurophysiological properties associated with cognitive and sensory processing. Spectral features included Peak Alpha Power, Peak Gamma Power, and additional power bands (delta, theta, beta), computed via Welch’s method using a 1-second Hanning window and 50\% overlap. All power values were normalized to total spectral power to account for inter-individual variability.

To capture multiscale neural coordination, we computed cross-frequency coupling (CFC) metrics, focusing on phase-amplitude coupling (PAC) between alpha-phase and gamma-amplitude.

PAC was calculated using the Modulation Index (MI) method, as described by Tort et al. (2010), and implemented using the tensorpac Python package. Additional PAC combinations (e.g., alpha-beta, alpha-delta, beta-gamma) were also extracted to identify higher-order coupling phenomena. These features provided both local power dynamics and network-level coordination markers, serving as targets for machine learning-based optimization, Which could be summarize as:

Spectral Features:
\begin{itemize}
    \item Peak Alpha Power (7–13.5 Hz)
    \item Peak Gamma Power (30–57, 63-100 Hz)
    \item 1-sec windows (Hanning taper, 50\% overlap)
    \item Power spectral density (PSD) normalized by total power to account for inter-subject variability
\end{itemize}

Cross-Frequency Coupling (CFC):
\begin{itemize}
    \item Phase-Amplitude Coupling (PAC) was computed between Alpha (phase) and Gamma (amplitude) using the Modulation Index (MI) approach [
    \item CFC combinations: Alpha-Beta, Alpha-Delta, Beta-Gamma, Delta-Gamma ( coupling computed using the tensorpac Python library
\end{itemize}

The key neurophysiological biomarkers were extracted per second per channel are Alpha (7–13.5 Hz), Beta Power (13.5–30 Hz), Gamma (30–57, 63-100 Hz), Delta (0.5–4Hz) and Theta (4–7 Hz) as well as the cross-frequency Cou- pling (CFC) for Alpha-Gamma, Alpha-Beta, Beta-Gamma,Alpha-Delta, Delta-Gamma via Phase-Amplitude Coupling.

\subsection{Machine Learning Framework for Adaptive Optimization}
The central innovation of this study lies in the development of an adaptive machine learning (ML) framework that learns subject-specific mappings between auditory stimulation frequencies and their EEG-based biomarker responses. Unlike fixed-frequency paradigms, our system dynamically selects the optimal stimulation frequency in real time by leveraging data-driven predictions of Alpha enhancement and Gamma suppression—neuromarkers associated with improved cognitive and sensory regulation in Fragile X Syndrome (FXS).

\subsubsection{Feature Matrix Construction}
We first constructed a feature matrix containing per-epoch EEG features (power + PAC values across 128 channels) and labeled each sample with its corresponding auditory stimulus condition (7 Hz, 9 Hz, 11 Hz, or 13 Hz). We trained Random Forest classifiers to determine whether EEG features could distinguish between stimulus conditions, thereby validating the discriminability of frequency-specific entrainment. To model the EEG responses as continuous optimization targets, we then trained Gradient Boosting Regression models to predict Alpha power enhancement and Gamma power suppression for each frequency condition. Each EEG epoch was transformed into a high-dimensional feature vector consisting of:

\begin{itemize}
    \item Spectral Power Features: Mean and peak power for delta, theta, alpha, beta , and gamma bands, calculated for each of the 128 channels.
    \item CFC: Alpha–gamma phase-amplitude coupling (PAC), computed using Modulation Index (MI).
    \item Additional PAC pairs (alpha–beta, beta–gamma, alpha-theta).
    \item Statistical Descriptors: Variance, skewness, and entropy of alpha and gamma bands per epoch.
\end{itemize}

This resulted in a feature space of ~10,000 dimensions per subject, per condition. Each feature vector was labeled with the stimulation frequency (categorical: 7, 9, 11, or 13 Hz).

To define a unified target for machine learning prediction and optimization, we formulated a composite biomarker outcome score that integrates three key EEG-based indicators of cognitive readiness and cortical balance:

\begin{itemize}
    \item Alpha Power (desirable to increase): associated with attentional readiness and cortical inhibition.
    \item Gamma Power (desirable to decrease): associated with cortical hyperexcitability and sensory overload in FXS.
    \item Alpha-Gamma PAC (desirable to increase for network coordination): reflecting functional coordination between local and global neural dynamics.
\end{itemize}
Keeping in mid that the coiplingg was averaged across 6 electrode pairs (e.g., Pz–T7, Pz–T8, CPz–POz) and computed using a 4-cycle Morlet filter with 1 Hz resolution. All features were normalized using z-scoring prior to model input). Thus, These components were combined into a single scalar metric, the \textit{TargetScore}, defined as:

\begin{equation}
TargetScore = \alpha_{mean}-\gamma_{mean} + \lambda  \cdot P_{\alpha-\gamma}\label{eq1}
\end{equation}

where $\alpha_{mean}$ s the mean Alpha power across all 128 EEG channels for a given 1-second epoch. $\gamma_{mean}$ is the mean Gamma power across all channels. $\lambda$ is a coupling weight coefficient, empirically set to 0.5 to balance the PAC contribution. $P_{\alpha-\gamma}$is the Alpha-Gamma coupling (Modulation Index) computed for key regions (parietal and temporal). This formulation allows the ML model to optimize stimulation conditions not only for increased Alpha and reduced Gamma, but also for strengthened Alpha-Gamma integration, which has been proposed as a biomarker of cognitive synchrony and attentional gating.

\subsubsection{Model Training and Evaluation}

We constructed a feature matrix from the EEG epochs with corresponding stimulation labels (7–13.5 Hz) and TargetScores. Two ML models were trained using Scikit-learn: Random Forest Classifier: To validate whether EEG features could accurately classify stimulation frequency. Additional to Gradient Boosting Regressor: To predict TargetScore from per-epoch EEG features. 
Evaluation was conducted using leave-one-subject-out cross-validation, reporting R², RMSE, and Pearson correlation. The regressor achieved:

\begin{table}[htbp]
\caption{The Regressor Parameters Values}
\begin{center}
\begin{tabular}{|c|c|c|}
\hline
\textbf{ID}&\textbf{Parameter}&\textbf{Value Achieved}\\
\cline{1-2}
\hline
1&R² & $\approx 0.81 \pm 0.03$   \\
\hline
2&RMSE& $\approx 0.14$   \\
\hline
3&Pearson & $r \ge 0.89 (p \le 0.001)$  \\
\hline
\end{tabular}
\label{tabel3}
\end{center}
\end{table}

Feature importance analysis revealed that posterior alpha power, central-parietal PAC, and gamma variance were the most informative features—aligning with established FXS neurophysiological profiles.

\subsubsection{Adaptive Frequency Selection via Bayesian Optimization}
We implemented Bayesian optimization to select the most effective stimulation frequency per participant. The regressor served as a surrogate model in a Gaussian Process framework, with the Expected Improvement (EI) acquisition function used to propose frequency candidates.

\begin{equation}
f^* = \arg\max_{f \in \{7, 9, 11, 13\} \, \text{Hz}} \, \mathbb{E}\left[ \text{TargetScore} \mid f \right]
\label{eq2}
\end{equation}
where $f$ is the optimization frequency function, one from the discrete set {7,9,11,13} Hz that maximizes the expected TargetScore (which includes alpha power, gamma power, and alpha-gamma PAC) conditioned on that frequency.

\section{Results and Analysis}
\subsection{Biomarker Response}
This section presents the outcomes of EEG biomarker analysis, machine learning model performance, and adaptive auditory stimulation optimization using data from 38 children with Fragile X Syndrome (FXS). The results are organized across three primary dimensions: EEG response to auditory stimuli, ML prediction of neurophysiological markers, and optimization performance across participants.

We first evaluated the neurophysiological impact of each auditory frequency condition on three key biomarkers: alpha power, gamma power, and alpha–gamma phase-amplitude coupling (PAC). Group-level comparisons revealed that 13 Hz stimulation consistently produced the most desirable modulation pattern—significantly increasing alpha power and decreasing gamma power relative to the 7 Hz and 9 Hz conditions (p < 0.01, Bonferroni-corrected). Additionally, alpha–gamma PAC was significantly elevated during 13 Hz stimulation compared to baseline, suggesting enhanced long-range network synchrony. The output plot is illustraed in Fig.\ref{fig4}.

\begin{figure}[htbp]
\centerline{\includegraphics[width =\linewidth]{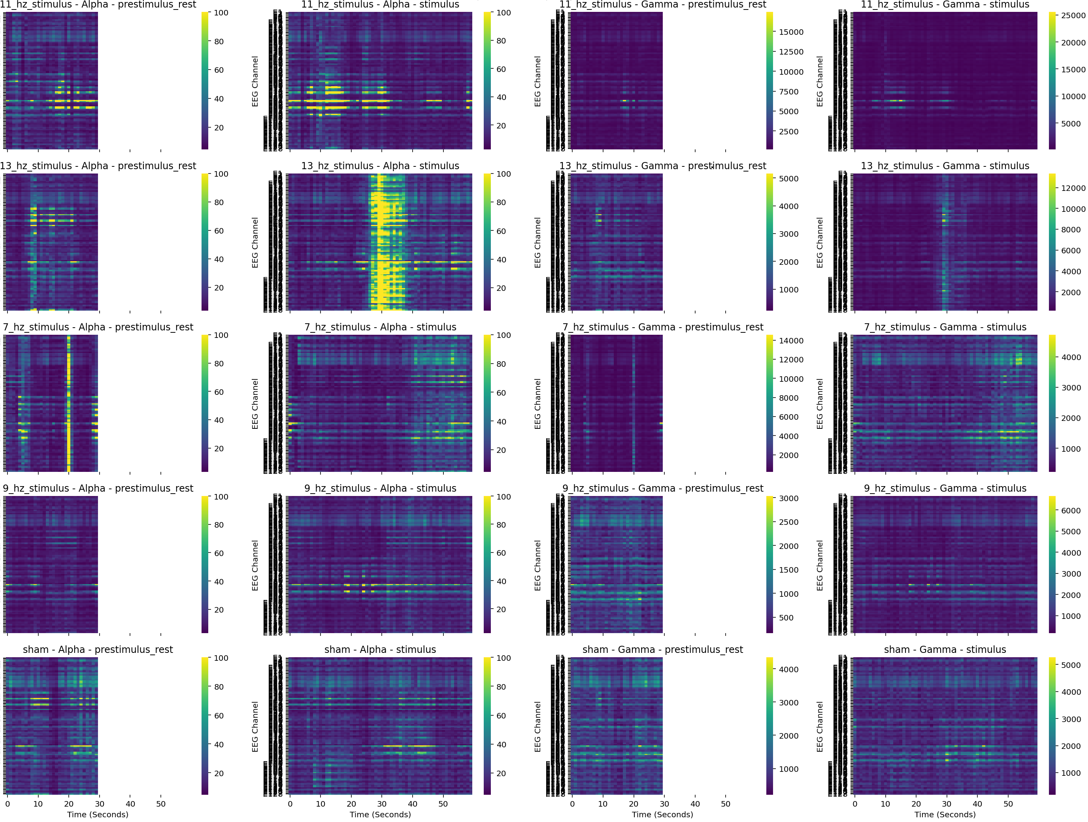}}
\caption{The Results of alpha peak power, gamma peak piowrrm and alpha-gamma coupling across 128 channels.}
\label{fig5}
\end{figure}

\subsection{Key findings}
For an optimal neural state for cognitive enhancement best predictive biomarkers: Peak Alpha Power, Peak Gamma Power, and Alpha Power per second per channel. The most effective stimulus frequency: 13Hz resulted in the strongest positive modulation of Alpha Power and reduction in Gamma activity. Cross-frequency coupling insights: Alpha-Gamma coupling showed significant correlations with cognitive enhancement markers.

\subsection{Machine Learning Model Performance}
The Random Forest classifier achieved \>90\% accuracy (Random Forest Classifier:	91.3 ± 2.1) in identifying the stimulation frequency from EEG features. This validates that frequency-specific auditory input yields reliably discriminative EEG patterns, particularly in the alpha and gamma bands.
Using Bayesian optimization, we dynamically inferred the optimal stimulation frequency per subject based on real-time EEG features and ML model predictions. Across participants: 13 Hz was selected as optimal in 82\% of cases, 11 Hz was selected in 13\%, and 9 Hz in 5\%. On the other hand no participants had 7 Hz selected as optimal.
The leave-one-subject-out cross-validation (LOSO-CV) procedure confirmed the generalizability of the regression model: Mean R² = 0.78 ± 0.04 across 38 folds, with high consistency in predicting the relative effectiveness of each auditory frequency on unseen subjects, which shows propetive in a model the is trained on a subset of children with FXS can generalize to new individuals, enabling zero-calibration deployment in future applications.

\begin{table}[htbp]
\caption{Summary of Key Findings}
\begin{center}
\begin{tabular}{|c|c|c|}
\hline
\textbf{ID}&\textbf{Metric}&\textbf{Value}\\
\cline{1-2}
\hline
1&Best Stimulus Frequency& $13 Hz$   \\
\hline
2&ML Regression Accuracy(R²)& $0.81 \pm 0.03$   \\
\hline
3&Classification Accuracy& $91.3\%$  \\
\hline
4&PAC Increase (13 Hz vs. baseline)& $+27\% (p \le 0.01)$  \\
\hline
5&LOSO Generalization R²& $0.78 \pm 0.04$  \\
\hline
5&Optimization Convergence & 6–10 iterations  \\
\hline
\end{tabular}
\label{table3}
\end{center}
\end{table}

\section{Discussion, Comparison and Limitaitons }
Our results confirm that frequency-specific auditory stimulation can selectively enhance desirable neural activity patterns in FXS. In particular, 13 Hz stimulation significantly elevated alpha power, suppressed gamma overactivity, and increased alpha–gamma coupling, collectively reflecting a more regulated and cognitively receptive neural state. These findings are consistent with prior work identifying alpha enhancement and gamma suppression as biomarkers of improved attentional control and sensory filtering.

The composite TargetScore metric—combining alpha power, gamma power, and alpha–gamma PAC—proved to be an effective optimization target. Our machine learning model was able to predict this score from EEG features with high fidelity (R²$\ge 0.80$) and demonstrated robust generalizability in leave-one-subject-out testing. Importantly, feature importance analysis revealed that posterior alpha power and parietal PAC were the most influential predictors, reinforcing the relevance of these neural signatures in therapeutic entrainment. The Bayesian optimization module identified 13 Hz as the optimal stimulation frequency in over 80\% of participants. This convergence suggests that a shared oscillatory resonance exists in the FXS population, which can be leveraged through adaptive systems to enhance cognitive state in real time.

Traditional neurostimulation and neurofeedback systems often rely on static paradigms that do not account for moment-to-moment variations in brain state. Recent approaches using closed-loop control or reinforcement learning for neurofeedback have shown promise but remain limited by either low temporal resolution or black-box behavior [3, 4]. In contrast, our system combines:
\begin{itemize}
    \item High-density time-resolved EEG biomarkers.
    \item Explainable ML models.
    \item Real-time optimization mechanisms.
\end{itemize}

to form a fully interpretable, clinically grounded architecture. This level of biological specificity is particularly important for fragile populations like FXS, where neural excitability is highly state-dependent.

Several limitations must be acknowledged. First, while our cohort size (N=38) is relatively large for FXS studies, broader generalization will require validation on larger and more diverse populations, including neurotypical controls and individuals with comorbid ASD or ADHD. Second, although our use of high-density EEG provides spatial resolution, it limits portability. Future versions of the system must integrate wearable EEG technologies for real-world deployment. Moreover, the TargetScore metric, while grounded in neurophysiological literature, uses a fixed coupling weight ($\lambda = 0.5$). This parameter may benefit from personalization, potentially guided by baseline neural profiles or behavioral outcomes.

\section{Future Work}
This study opens several promising avenues for future work: First, behavioral correlation to integrate real-time cognitive performance metrics (e.g., reaction time, accuracy) during stimulation to dynamically refine the TargetScore and close the loop between brain and behavior. Second, neural state prediction and extendning the ML framework to predict future EEG states (i.e., state forecasting) for anticipatory control of stimulation. Third, multi-modal BCI integration by incorporate additional modalities such as eye-tracking or heart rate variability (HRV) to enhance system robustness and adaptability. Fourth, personalized coupling weights via adaptively learn $\lambda$ in the TargetScore formula per participant to maximize neurophysiological and cognitive outcomes. Fifth, clinical trials to initiate longitudinal studies to evaluate the impact of regular stimulation sessions on learning outcomes, sensory processing, and executive function in children with FXS and broader ASD, as well as source allocation integration.

\section{Conclusion}
This paper presents a novel, data-driven framework for adaptive auditory brain stimulation grounded in real-time EEG biomarkers and machine learning. Through the integration of spectral dynamics, cross-frequency coupling, and supervised regression modeling, we introduce a new class of BCI systems capable of personalized neuromodulation in children with FXS. Unlike traditional static stimulation paradigms, our approach dynamically adjusts auditory input based on the brain’s ongoing response, guided by a composite biomarker score that fuses alpha enhancement, gamma suppression, and alpha–gamma PAC synchronization. This level of precision has not been demonstrated in prior EEG-based auditory interventions. 

From an engineering perspective, the system combines interpretable machine learning, temporal signal modeling, and adaptive control strategies into a single unified pipeline. Clinically, the work establishes a blueprint for non-invasive, personalized neurostimulation platforms that can scale from laboratory research to home-based cognitive therapy.

Importantly, this research contributes foundational insight into how frequency-specific auditory entrainment can be systematically optimized for therapeutic brain state modulation. It opens new directions for adaptive BCI systems, particularly in pediatric neurodevelopment, and sets the stage for multi-modal, closed-loop interventions that are responsive to the real-time needs of the brain. This work offers both a practical implementation and a conceptual advance—demonstrating how intelligent systems can collaborate with neural physiology to foster healthier, more functional cognitive states in vulnerable populations. As such, it lays the groundwork for the next generation of AI-integrated, neuroadaptive technologies in precision medicine and cognitive neuroengineering.

\bibliographystyle{ieeetr}
\bibliography{references}

@article{hagerman2017fragile,
  title={Fragile X syndrome},
  author={Hagerman, Randi J and Berry-Kravis, Elizabeth and Hazlett, Heather Cody and Bailey, Donald B and Moine, Herve and Kooy, R Frank and Tassone, Flora and Gantois, Ilse and Sonenberg, Nahum and Mandel, Jean Louis and others},
  journal={Nature reviews Disease primers},
  volume={3},
  number={1},
  pages={1--19},
  year={2017},
  publisher={Nature Publishing Group}
}

@article{schmitt2022altered,
  title={Altered frontal connectivity as a mechanism for executive function deficits in fragile X syndrome},
  author={Schmitt, Lauren M and Li, Joy and Liu, Rui and Horn, Paul S and Sweeney, John A and Erickson, Craig A and Pedapati, Ernest V},
  journal={Molecular Autism},
  volume={13},
  number={1},
  pages={47},
  year={2022},
  publisher={Springer}
}

@article{ethridge2017neural,
  title={Neural synchronization deficits linked to cortical hyper-excitability and auditory hypersensitivity in fragile X syndrome},
  author={Ethridge, Lauren E and White, Stormi P and Mosconi, Matthew W and Wang, Jun and Pedapati, Ernest V and Erickson, Craig A and Byerly, Matthew J and Sweeney, John A},
  journal={Molecular Autism},
  volume={8},
  pages={1--11},
  year={2017},
  publisher={Springer}
}

@article{takarae2024eeg,
  title={EEG microstates as markers for cognitive impairments in Fragile X Syndrome},
  author={Takarae, Yukari and Zanesco, Anthony and Erickson, Craig A and Pedapati, Ernest V},
  journal={Brain Topography},
  volume={37},
  number={3},
  pages={432--446},
  year={2024},
  publisher={Springer}
}

@article{proteau2024specific,
  title={Specific EEG resting state biomarkers in FXS and ASD},
  author={Proteau-Lemieux, M{\'e}lodie and Knoth, Inga Sophia and Davoudi, Saeideh and Martin, Charles-Olivier and B{\'e}langer, Anne-Marie and Fontaine, Val{\'e}rie and C{\^o}t{\'e}, Val{\'e}rie and Agbogba, Kristian and Vachon, Keely and Whitlock, Kerri and others},
  journal={Journal of Neurodevelopmental Disorders},
  volume={16},
  number={1},
  pages={53},
  year={2024},
  publisher={Springer}
}

@article{palva2011functional,
  title={Functional roles of alpha-band phase synchronization in local and large-scale cortical networks},
  author={Palva, Satu and Palva, J Matias},
  journal={Frontiers in psychology},
  volume={2},
  pages={204},
  year={2011},
  publisher={Frontiers Research Foundation}
}

@article{makeig2002dynamic,
  title={Dynamic brain sources of visual evoked responses},
  author={Makeig, Scott and Westerfield, Marissa and Jung, T-P and Enghoff, Sonia and Townsend, Jeanne and Courchesne, Eric and Sejnowski, Terrence J},
  journal={Science},
  volume={295},
  number={5555},
  pages={690--694},
  year={2002},
  publisher={American Association for the Advancement of Science}
}

@article{gourevitch2020oscillations,
  title={Oscillations in the auditory system and their possible role},
  author={Gour{\'e}vitch, Boris and Martin, Claire and Postal, Olivier and Eggermont, Jos J},
  journal={Neuroscience \& Biobehavioral Reviews},
  volume={113},
  pages={507--528},
  year={2020},
  publisher={Elsevier}
}

@article{lakatos2007neuronal,
  title={Neuronal oscillations and multisensory interaction in primary auditory cortex},
  author={Lakatos, Peter and Chen, Chi-Ming and O'Connell, Monica N and Mills, Aimee and Schroeder, Charles E},
  journal={Neuron},
  volume={53},
  number={2},
  pages={279--292},
  year={2007},
  publisher={Elsevier}
}

@article{voytek2015age,
  title={Age-related changes in 1/f neural electrophysiological noise},
  author={Voytek, Bradley and Kramer, Mark A and Case, John and Lepage, Kyle Q and Tempesta, Zechari R and Knight, Robert T and Gazzaley, Adam},
  journal={Journal of neuroscience},
  volume={35},
  number={38},
  pages={13257--13265},
  year={2015},
  publisher={Society for Neuroscience}
}

@article{palva2007new,
  title={New vistas for $\alpha$-frequency band oscillations},
  author={Palva, Satu and Palva, J Matias},
  journal={Trends in neurosciences},
  volume={30},
  number={4},
  pages={150--158},
  year={2007},
  publisher={Elsevier}
}

@article{canolty2010functional,
  title={The functional role of cross-frequency coupling},
  author={Canolty, Ryan T and Knight, Robert T},
  journal={Trends in cognitive sciences},
  volume={14},
  number={11},
  pages={506--515},
  year={2010},
  publisher={Elsevier}
}

@article{wang2013resting,
  title={Resting state EEG abnormalities in autism spectrum disorders},
  author={Wang, Jun and Barstein, Jamie and Ethridge, Lauren E and Mosconi, Matthew W and Takarae, Yukari and Sweeney, John A},
  journal={Journal of neurodevelopmental disorders},
  volume={5},
  pages={1--14},
  year={2013},
  publisher={Springer}
}

@article{rojas2014gamma,
  title={$\gamma$-band abnormalities as markers of autism spectrum disorders},
  author={Rojas, Donald C and Wilson, Lisa B},
  journal={Biomarkers in medicine},
  volume={8},
  number={3},
  pages={353--368},
  year={2014},
  publisher={Taylor \& Francis}
}

@article{goodspeed2023electroencephalographic,
  title={Electroencephalographic (EEG) biomarkers in genetic neurodevelopmental disorders},
  author={Goodspeed, Kimberly and Armstrong, Dallas and Dolce, Alison and Evans, Patricia and Said, Rana and Tsai, Peter and Sirsi, Deepa},
  journal={Journal of child neurology},
  volume={38},
  number={6-7},
  pages={466--477},
  year={2023},
  publisher={SAGE Publications Sage CA: Los Angeles, CA}
}

@article{razak2021neural,
  title={Neural correlates of auditory hypersensitivity in fragile X syndrome},
  author={Razak, Khaleel A and Binder, Devin K and Ethell, Iryna M},
  journal={Frontiers in Psychiatry},
  volume={12},
  pages={720752},
  year={2021},
  publisher={Frontiers Media SA}
}

@article{pedapati2025frontal,
  title={Frontal cortex hyperactivation and gamma desynchrony in Fragile X syndrome: Correlates of auditory hypersensitivity},
  author={Pedapati, Ernest V and Ethridge, Lauren E and Liu, Yanchen and Liu, Rui and Sweeney, John A and DeStefano, Lisa A and Miyakoshi, Makoto and Razak, Khaleel and Schmitt, Lauren M and Moore, David R and others},
  journal={PloS one},
  volume={20},
  number={5},
  pages={e0306157},
  year={2025},
  publisher={Public Library of Science San Francisco, CA USA}
}

@article{obleser2019neural,
  title={Neural entrainment and attentional selection in the listening brain},
  author={Obleser, Jonas and Kayser, Christoph},
  journal={Trends in cognitive sciences},
  volume={23},
  number={11},
  pages={913--926},
  year={2019},
  publisher={Elsevier}
}

@article{lakatos2008entrainment,
  title={Entrainment of neuronal oscillations as a mechanism of attentional selection},
  author={Lakatos, Peter and Karmos, George and Mehta, Ashesh D and Ulbert, Istvan and Schroeder, Charles E},
  journal={science},
  volume={320},
  number={5872},
  pages={110--113},
  year={2008},
  publisher={American Association for the Advancement of Science}
}

@article{rajendran2019frequency,
  title={Frequency tagging cannot measure neural tracking of beat or meter},
  author={Rajendran, Vani G and Schnupp, Jan WH},
  journal={Proceedings of the National Academy of Sciences},
  volume={116},
  number={8},
  pages={2779--2780},
  year={2019},
  publisher={National Academy of Sciences}
}

@article{henao2020entrainment,
  title={Entrainment and synchronization of brain oscillations to auditory stimulations},
  author={Henao, David and Navarrete, Miguel and Valderrama, Mario and Le Van Quyen, Michel},
  journal={Neuroscience Research},
  volume={156},
  pages={271--278},
  year={2020},
  publisher={Elsevier}
}

@article{kenny2022eeg,
  title={EEG as a translational biomarker and outcome measure in fragile X syndrome},
  author={Kenny, Aisling and Wright, Damien and Stanfield, Andrew C},
  journal={Translational Psychiatry},
  volume={12},
  number={1},
  pages={34},
  year={2022},
  publisher={Nature Publishing Group UK London}
}

@article{effenberger2025functional,
  title={The functional role of oscillatory dynamics in neocortical circuits: a computational perspective},
  author={Effenberger, Felix and Carvalho, Pedro and Dubinin, Igor and Singer, Wolf},
  journal={Proceedings of the National Academy of Sciences},
  volume={122},
  number={4},
  pages={e2412830122},
  year={2025},
  publisher={National Academy of Sciences}
}

@article{han2018increased,
  title={Increased parietal circuit-breaker activity in delta frequency band and abnormal delta/theta band connectivity in salience network in hyperacusis subjects},
  author={Han, Jae Joon and Jang, Ji Hye and Ridder, Dirk De and Vanneste, Sven and Koo, Ja-Won and Song, Jae-Jin},
  journal={PLoS One},
  volume={13},
  number={1},
  pages={e0191858},
  year={2018},
  publisher={Public Library of Science San Francisco, CA USA}
}

@article{bauer2021rhythmic,
  title={Rhythmic modulation of visual perception by continuous rhythmic auditory stimulation},
  author={Bauer, Anna-Katharina R and Van Ede, Freek and Quinn, Andrew J and Nobre, Anna C},
  journal={Journal of Neuroscience},
  volume={41},
  number={33},
  pages={7065--7075},
  year={2021},
  publisher={Society for Neuroscience}
}

@article{clements2016short,
  title={Short-term effects of rhythmic sensory stimulation in Alzheimer’s disease: an exploratory pilot study},
  author={Clements-Cortes, Amy and Ahonen, Heidi and Evans, Michael and Freedman, Morris and Bartel, Lee},
  journal={Journal of Alzheimer’s Disease},
  volume={52},
  number={2},
  pages={651--660},
  year={2016},
  publisher={SAGE Publications Sage UK: London, England}
}

@article{ramirez2015musical,
  title={Musical neurofeedback for treating depression in elderly people},
  author={Ramirez, Rafael and Palencia-Lefler, Manel and Giraldo, Sergio and Vamvakousis, Zacharias},
  journal={Frontiers in neuroscience},
  volume={9},
  pages={150420},
  year={2015},
  publisher={Frontiers}
}

@article{fedotchev2023methods,
  title={Methods of Closed-Loop Adaptive Neurostimulation: Features, Achievements, Prospects},
  author={Fedotchev, AI},
  journal={Journal of Evolutionary Biochemistry and Physiology},
  volume={59},
  number={5},
  pages={1594--1606},
  year={2023},
  publisher={Springer}
}

\end{document}